\documentclass[aps,prd,twocolumn,nofootinbib]{revtex4-1}
\usepackage[T1]{fontenc}
\usepackage[utf8]{inputenc}
\usepackage[english]{babel}
\usepackage{url}
\usepackage{mathrsfs}
\usepackage{amsfonts}
\usepackage{amsmath}
\usepackage{amssymb}
\usepackage{amsthm}
\usepackage{graphicx}

\usepackage[toc,title]{appendix}
\usepackage{bm}
\usepackage{booktabs}
\usepackage{braket}
\usepackage{caption}
\usepackage{enumerate}
\usepackage{enumitem}
\usepackage[Bjornstrup]{fncychap} 
\usepackage{fnpos} %
\usepackage[bottom,multiple]{footmisc} 
\usepackage{lipsum}	
\usepackage{mathtools} 
\usepackage[]{natbib}	
\usepackage{quoting}
\quotingsetup{font=small}
\usepackage{sidecap}
\usepackage{subfig}
\usepackage{textcomp}
\usepackage{tikz}
\usetikzlibrary{decorations.pathmorphing,calc}
\usepackage{titlesec}
\usepackage{tensor}
\usepackage{xfrac}

\usepackage{csquotes}
\usepackage{helvet}
\usepackage{listings}

\usepackage[bookmarks,bookmarksopen=true,colorlinks,citecolor=blue,linkcolor=blue,urlcolor=blue]{hyperref}

\makeFNbelow
\newcommand{\be}{\begin{equation}}
\newcommand{\ee}{\end{equation}}
\newcommand{\bn}{\begin{equation*}}
\newcommand{\en}{\end{equation*}}

\newcommand{\cit}{\citep}		

\newcommand{\lie}{\pounds} 
\newcommand{\Ae}{\text{\AE}} 
\newcommand{\ux}{(u\cdot \xi)}
\newcommand{\sx}{(s\cdot \xi)}
\newcommand{\ut}{(u\cdot t)}
\newcommand{\st}{(s\cdot t)}
\newcommand{\as}{(a\cdot s)}

\newcommand{\M}{\mathcal{M}}
\newcommand{\J}{\mathcal{J}}
\newcommand{\ru}{r_{\text{UH}}}
\newcommand{\rk}{r_{\text{KH}}}
\newcommand{\ku}{\kappa_{\text{UH}}}
\newcommand{\kk}{\kappa_{\text{KH}}}
\newcommand{\f}{\varphi} 

\DeclarePairedDelimiter{\abs}{\lvert}{\rvert}	

\theoremstyle{plain}

\begin{document}

\rmfamily

\title{An improved derivation of the Smarr Formula\\ for Lorentz--breaking gravity}
\date{\today}
\author{Costantino Pacilio}
\author{Stefano Liberati}
\affiliation{SISSA, Via Bonomea 265, 34136 Trieste, Italy, EU}
\affiliation{INFN, Sezione di Trieste}
\begin{abstract}
Thermodynamical properties of black holes in gravitational theories without Local Lorentz invariance have been subject to intense investigation in the past years due to the presence of universal horizons, which are strong causal barriers even for superluminal signals.  Here we present a novel general method for deriving the Smarr formula for this class of theories: in particular we show that the Smarr formulae for Einstein--{\Ae}ther theory and infrared Ho{\v r}ava gravity follow from scale invariance. We not only reproduce straightforwardly previous findings for static black hole solutions, but we are also able to generalise them to the case of stationary rotating black holes. Finally, we apply our results to the rotating black holes with universal horizon as found in three dimensions, from which we shall draw some lessons on the viability of black hole thermodynamics for black hole solutions endowed with universal horizons.
\end{abstract}
\maketitle
%
\section{Introduction}

Revived by the overwhelming pressure of cosmological observations --- with their still puzzling dark sectors required within General Relativity (GR) to fit the data --- modified theories of gravity have received a growing attention in the last twenty years~\cite{Clifton:2011jh}. Among these, theories where Local Lorentz Symmetry (LLS) is violated have been enjoying a growing interest also for they relevance for quantum gravity as well as as a test bed for our understanding of black hole thermodynamics.

The easiest way to break LLS, while preserving general covariance, is to introduce a dynamical preferred timelike vector field. This is the strategy of Einstein--{\Ae}ther (\Ae) theory \cite{jacobson_report}, a vector--tensor theory of gravitation. The theory is conceived as an effective modification of GR, where only second order time derivatives in the EOM are admitted. Ho{\v r}ava gravity \cite{Horava:2009uw} instead introduces a preferred timelike foliation, which can be made dynamical \cite{blas_extension, Horava_status, wang_review}. The aim is to provide an UV completion of GR, in such a way that the resulting theory is renormalizable. Indeed, power counting renormalizability of a subclass of the theory was recently proved in~\cite{Barvinsky:2015kil}.

In \cite{jacobson_Horava, jacobson_undoing} it was shown that \Ae--theory is closely related to the infrared limit of Ho{\v r}ava gravity, obtained by neglecting more-than-second order operators in the Ho{\v r}ava action. In particular, restricting the {\Ae} field to be hypersurface orthogonal at the level of the action, the resulting theory is equivalent to infrared Ho{\v r}ava gravity. Moreover in \cite{Horava_lower} a formal algorithm to construct an \Ae\, analog of the full Ho{\v r}ava theory was also proposed.

The physics of stars and black holes represents an important arena where to test these theories, both observationally and theoretically. Astrophysical constraints for compact objects were studied in \cite[see also \citenum{jacobson_report,Horava_status,liberati_lorentz}]{binary_1, binary_2, binary_3}. At the same time theoretical aspects have been investigated, and in particular it became clear that Lorentz violation (LV) produces also novel causal structures \cite{barausse_black, blas_uh}. More specifically, a generalization of the concept of black hole emerges, in which the usual Killing horizon coexists with a new type of horizon, called universal horizon (UH). The UH is an horizon which traps modes of arbitrary speed which are indeed allowed in these theories. Therefore the existence of the UH provides a meaningful notion of BH when LLS is broken. (For an extensive treatment of causal structure in presence of Lorentz violation, and a formal treatment of Universal Horizons, see \cite{sotiriou_uh_global}.)

Among several theoretical aspects, much attention was given to the issue of BH mechanics and thermodynamics. Doubts that Killing Horizons have a meaningful BH mechanics in LV theories were raised in~\cite{foster_noether}. On the contrary, indications that universal horizons can have a well defined mechanics came from~\cite{mattingly_mechanics, bhatta_thesis}: here two exact 4--dimensional static asymptotically flat BH solutions endowed with universal horizon were found, and the corresponding Smarr Formulas and First Laws were computed.

Attempts to provide a thermodynamical interpretation of the UH mechanics were pursued in~\cite{mattingly_tunneling, bhatta_thesis, ding_radiation} where a temperature associated to such horizons was computed via a tunnelling approach. The connection has been further explored in \cite{cropp_ray, cropp_thesis}, where the UH tunnelling temperature was related to an appropriate notion of surface gravity. However contrasting results also exist~\cite{parentani}, claiming that the LLS breaking black holes still radiate with the usual temperature associated to the Killing Horizon, as in the Lorentz symmetric case.

All these analyses have been limited so far to static BH solutions. Staticity is a high degree of symmetry and can be a source of degeneracy, therefore the study of rotating solutions would represent a potential resource to remove ambiguities. Unfortunately no 4-dimensional fully rotating BH solutions have been found so far, and therefore this analysis is still precluded.

In this paper we make a step towards the extension of LV BH mechanics beyond the static case. We present a method for deriving a Smarr Formula in \Ae--theory and IR Ho{\v r}ava gravity, which relies on the fact that these theories are scale invariant. The advantage of this new derivation is that it streamlines and simplifies the previous one \cite{mattingly_mechanics, bhatta_thesis}, based upon staticity, and it can be applied to more general solutions, including rotating BHs. As an application, we compute the Smarr Formula of the 3--dimensional fully rotating solutions discovered in \cite{btz_uh}.

We organize the exposition as follows. Sec.\,\ref{sec:ae:review} reviews \Ae--theory and fixes the notations and conventions. In Sec.\,\ref{subsec:derivation} we derive the generalized Smarr Formula, and in Sec.\,\ref{subsec:scale} we clarify the role of scale invariance and we comment about the inclusion of the cosmological constant. Sec.\,\ref{subsec:static} shows that our Smarr Formula reproduces the one in \cite{mattingly_mechanics}, when restricted to the static case. In Sec.\,\ref{sec:Ho{\v r}ava} we discuss the extension to the IR Ho{\v r}ava gravity. Sec.\,\ref{sec:rotating} contanins an application of the above results to the case of the rotating 3--dimensional BHs of \cite{btz_uh}. Finally Sec.\,\ref{sec:conclusion} contains an overview of the results and concluding remarks.

We use the conventions adapted to the mostly plus metric signature $(-+++)$. Moreover, in sections \ref{sec:ae:review}--\ref{sec:Ho{\v r}ava}, we work in generic $D>2$ spacetime dimensions.

\section{Review of \AE--theory}
\label{sec:ae:review}

\AE--theory \cite{jacobson_report} is a generally covariant modification of GR, in which a dynamical four-vector field $u^a$ (the aether) is present, in addition to the usual metric tensor $g_{ab}$. The aether $u^a$ is constrained to be unit-timelike, and therefore its vacuum configuration defines a preferred timelike direction, thus breaking the original local Lorentz symmetry present at the level of the action. The theory is intended to be an effective description of possible Lorentz-violating physics in the gravity sector. In this spirit, one writes the \AE\, action as the one containing all the possible terms giving rise to second order field equations~\cite{jacobson_report, eling_static}:
\be
\label{eq:ae:action}
S_\Ae=\frac{1}{2\varkappa}\int d^Dx\sqrt{-g}\left[R+L_u\right],\qquad \varkappa=8\pi G
\ee
where
\be
\label{eq:ae:lu}
L_u=-K\indices{^{ab}_{cd}}\nabla_au^c\nabla_bu^d
\ee
and
\be
\label{eq:ae:k}
K\indices{^{ab}_{cd}}=c_1g^{ab}g_{cd}+c_2\delta^a_c\delta^b_d+c_3\delta^a_d\delta^b_c-c_4u^au^bg_{cd}.
\ee

The \AE\, Lagrangian is more transparent when expressed in the fluid form introduced in \cite{jacobson_undoing}. One can define the expansion, the shear and the twist of the aether field by using congruences of its flow--lines:
\begin{subequations}
\begin{align}
&\theta=\nabla\cdot u&&\text{expansion},\\
&\sigma_{ab}=\underleftarrow{\nabla_{(a}u_{b)}}&&\text{shear},\\
&\omega_{ab}=\underleftarrow{\nabla_{[a}u_{b]}}&&\text{twist},
\end{align}
\end{subequations}
where the under-left arrow indicates projection of the indices on the tangent space orthogonal to $u^a$.

Then the Lagrangian $L_u$ reads
\be
L_u=-\left[\frac{c_\theta}{D-1}\theta^2+c_\sigma\abs{\sigma}^2+c_\omega\abs{\omega}^2-c_aa^2\right],
\ee
and the relation between the fluid coefficients and the usual ones is
\begin{subequations}
\label{eq:fluid}
\begin{align}
&c_\theta=(D-1)c_2+c_{13},&&c_\sigma=c_{13},\\
& c_\omega=c_1-c_3,&&c_a=c_{14},
\end{align}
\end{subequations}
where we use the notation $c_{13}=c_1+c_3$, etc.

In order to impose the unit--timelike constraint on $u^a$, the action \eqref{eq:ae:action} is supplied with a Lagrange multiplier, and the total action becomes
\be
\label{eq:ae:action2}
S=S_\Ae+\frac{1}{2\varkappa}\int d^Dx\sqrt{-g}\,\lambda\left(u_au_bg^{ab}+1\right).
\ee
In this paper we will make use of the covariant symplectic formalism for diffeoinvariant theories, as it was carried out in \cite{lee_wald, waldentropy1, waldentropy2}. Recall that, under a generic variation $\delta\f$ of the dynamical fields $\{\f\}$, the variation of the Lagrangian D-form $\mathbb{L}$  can be expressed as a sum of a bulk term plus a boundary one:
\be
\label{eq:varl}
\delta\mathbb{L}=\mathbb{E}_{\f}\delta\f+d\Theta(\f,\delta\f)
\ee
where the "symplectic potential" $\Theta$ is a $(D-1)$-form locally constructed out of $\f$ and $\delta\f$, and it's linear and homogeneous in $\delta\f$. From \eqref{eq:varl} we read that the EOM corrsponding to each $\f$ are $\mathbb{E}_{\f}=0$.

When we vary the action \eqref{eq:ae:action2}, there is an ambiguity about which field to consider as independent: we choose to vary w.r.t. $g^{ab}$ and $u^a$. The corresponding variation of $S$ is
\begin{multline}
\label{eq:ae:vars}
\delta S=\frac{1}{2\varkappa}\int d^Dx\sqrt{-g}\left[\delta g^{ab}E_{ab}+2\delta u^a\left(\Ae_a+\lambda u_a\right)\right.\\
\left.+\delta\lambda(u^2+1)\right] +d\Theta
\end{multline}
where 
\be
\Ae_a=\frac{1}{2}\frac{\delta L_u}{\delta u^a}=c_4\,a^m\nabla_au_m+\nabla_mY\indices{^m_a},
\ee
$a^a=u^b\nabla_bu^a$ is the aether acceleration, and $\Theta$ and $E_{ab}$ are explicitated later.

Defining the two tensors
\begin{subequations}
\label{eq:ae:yx}
\begin{align}
&Y\indices{^a_b}=K\indices{^{ac}_{bd}}\nabla_cu^d,\\
&X\indices{^m_{ab}}=u^mY_{(ab)}+u_{(a}Y\indices{^m_{b)}}-u_{(b}Y\indices{_{a)}^m},
\end{align}
\end{subequations}
the resulting EOM are
\begin{subequations}
\begin{align}
&u_au_b\,g^{ab}=-1, \label{eq:ae:eom:1}\\
&\Ae_a+\lambda u_a=0, \label{eq:ae:eom:2}\\
&E_{ab}=G_{ab}-T^u_{ab}=0, \label{eq:ae:eom:3}
\end{align}
\end{subequations}
where $G_{ab}$ is the Einstein tensor and $T^u_{ab}$ is the aether stress--energy tensor:
\be
\label{eq:ae:t}
\begin{split}
T^u_{ab}=&c_1\left(\nabla_au_m\nabla_bu^m-\nabla_mu_a\nabla^mu_b\right)+c_4a_aa_b+\\
&+\nabla_mX\indices{^m_{ab}}+\lambda u_au_b+\frac{1}{2}L_u\,g_{ab}.
\end{split}
\ee
Solving Eq. \eqref{eq:ae:eom:2} for $\lambda$, we obtain 
\be
\label{eq:ae:lambda}
\lambda=u\cdot\Ae=c_4a^2+u^a\nabla_bY\indices{^b_a}
\ee
which can be finally replaced in Eq.s \eqref{eq:ae:t}, in order to express the EOM in terms of $g^{ab}$ and $u^a$ only:
\begin{subequations}
\begin{align}
&u_au_bg^{ab}=-1 ,\label{eq:ae:eom:4}\\
&(\delta^b_a+u^bu_a)\Ae_b=\underleftarrow{\Ae}_a=0 ,\label{eq:ae:eom:5}\\
\begin{split}
&G_{ab}=c_1(\nabla_au_m\nabla_bu^m-\nabla_mu_a\nabla^mu_b)+c_4a_aa_b\\
&+\nabla_mX\indices{^m_{ab}}+(u\cdot\Ae) u_au_b+\frac{1}{2}L_u\,g_{ab}.
\end{split}\label{eq:ae:eom:6}
\end{align}
\end{subequations}
The covariant symplectic analysis of \Ae--theory was carried out in \cite{foster_noether, mohd_aether}. The form $\Theta$ turns out to be
\begin{multline}
\label{eq:ae:theta}
\Theta(\f,\delta\f)=\frac{1}{2\varkappa}\left[g_{ab}\nabla^m\delta g^{ab}-\nabla_a\delta g^{ma}+X\indices{^m_{ab}}\delta g^{ab}\right.\\
\left.-2Y\indices{^m_a}\delta u^a\right]\epsilon_m.
\end{multline}
Here, and in the rest of the paper, we use the following notation from \cite{jacobson_entropy}. Given a $(D-n)$-dimensional submanifold of the starting $D$-dimensional manifold, the symbol $\epsilon_{a_1\dots a_n}$ denotes the tensor-valued $(D-n)$-form defined as $\epsilon_{a_1\dots a_n}=\hat{\epsilon}_{a_1\dots a_n}\bar{\epsilon}$, where $\hat{\epsilon}_{a_1\dots a_n}$ is the $n$-normal to the submanifold, and $\bar{\epsilon}$ is the induced volume form of the submanifold. So, in particular, $\epsilon$ denotes the volume form of the entire $D$-dimensional manifold, while $\epsilon_m$ in \eqref{eq:ae:theta} is the vector-valued volume form of a given hypersurface. 
%
\section{The Smarr Formula for \Ae--theory}
\label{sec:smarr:review}
\subsection{Derivation}
\label{subsec:derivation}

As observed in \cite{waldentropy1,waldentropy2}, diffeoinvariant theories have a conserved Noether current $(D-1)$--form $\mathbb{J}[\xi]$, associated to the invariance w.r.t. any arbitrary vector field $\xi$, 
\be
\label{eq:j}
\mathbb{J}[\xi]=\Theta(\f,\lie_\xi\f)-\xi\cdot\mathbb{L}.
\ee
$\mathbb{J}[\xi]$ is conserved on shell\footnote{We use an upper dot to indicate that an equality holds only on shell, when $\mathbb{E}_\f=0$.}:
\be
d\mathbb{J}[\xi]=d\Theta(\f,\lie_\xi\f)-d\xi\cdot\mathbb{L}=-\mathbb{E}_\f\lie_\xi\f\doteq 0,
\ee
and the conservation of $\mathbb{J}[\xi]$ implies the existence of a $(D-2)$-form $\mathbb{Q}[\xi]$ \citep{wald1990, waldentropy1}
\be
\mathbb{J}[\xi]\doteq d\mathbb{Q}[\xi]\, ,
\label{eq:NC}
\ee
called the ``Noether charge'' associated to $\xi$. For \Ae--theory $\mathbb{Q}[\xi]$ is \cite{foster_noether, mohd_aether}
\be
\label{eq:ae:q}
\mathbb{Q}[\xi]=\frac{-1}{2\varkappa}\left[\nabla^a\xi^b+u^aY\indices{^b_c}\xi^c+u^aY\indices{_c^b}\xi^c+Y\indices{^{ab}}\ux\right]\epsilon_{ab}.
\ee

Now assume to pick up a solution $\{\f\}$ of the theory which possesses a dynamical symmetry $\xi$, meaning that there exists a vector field $\xi$ such that $\lie_\xi\f\doteq 0$. It then follows, from linearity and homogeneity w.r.t. $\delta\f$, that $\Theta(\f,\lie_\xi\f)\doteq 0$ and, consequently, $\mathbb{J}[\xi]+\xi\cdot\mathbb{L}\doteq 0$. Integrating this last expression over an hypersurface $\Sigma$ with boundary $\partial\Sigma$ and using \eqref{eq:NC}, one then gets
\be
\label{eq:smarr:id}
0\doteq\int_{\partial\Sigma}\mathbb{Q}[\xi]+\int_\Sigma\xi\cdot\mathbb{L}.
\ee

In \cit{pacilio_smarr} we observed that this equation can be used to derive the Smarr Formula for black holes, if one chooses $\partial\Sigma=S_\infty\cup S_{\text{BH}}$, and provided that the second integral can be turned into a surface integral. We are going to show that this last condition is always true, independently from the solution, in the case of \AE--theory.

The key point is the observation that the \AE\, Lagrangian is a total divergence \emph{on shell}, i.e.~when the EOM are satisfied. This can be easily shown by tracing Eq. \eqref{eq:ae:eom:6}:
\be
\begin{split}
\left(\frac{2-D}{2}\right)R&=c_4a^2+\nabla_m(g^{ab}X^m_{ab})-(u\cdot\Ae)+\frac{D}{2}L_u\\
&=\nabla_m(g^{ab}X^m_{ab})-u^a\nabla_mY\indices{^m_a}+\frac{D}{2}L_u\\
&=\nabla_m(g^{ab}X^m_{ab}-u^aY\indices{^m_a})+Y\indices{^m_a}\nabla_mu^a+\frac{D}{2}L_u\\
&=\nabla_m(g^{ab}X^m_{ab}-u^aY\indices{^m_a})+\left(\frac{D-2}{2}\right)L_u,
\end{split}
\ee
from which
\be
\label{eq:ae:Lshell}
\begin{split}
R+L_u&=\left(\frac{2}{D-2}\right)\nabla_m(Y\indices{^m_a}u^a-g^{ab}X\indices{^m_{ab}})\\
&=\left(\frac{2}{D-2}\right)\nabla_m(u^aY\indices{_a^m}-u^mY\indices{^a_a}).
\end{split}
\ee
Taking into account that the total Lagrangian is $L=R+L_u+\lambda(u^2-1)$, and that the latter term vanishes on shell, we have that the Lagrangian $D$--form is a total divergence on shell, $\mathbb{L}\doteq d\mathbb{A}$, with
\be
\label{eq:ae:A}
\mathbb{A}=\left(\frac{2}{D-2}\right)\frac{(u^aY\indices{_a^m}-u^mY\indices{^a_a})\epsilon_m}{2\varkappa}.
\ee
Now, since the fields are assumed to be invariant under the flow of $\xi$, we have $\xi\cdot\mathbb{L}=\xi\cdot d\mathbb{A}=\lie_\xi\mathbb{A}-d(\xi\cdot\mathbb{A})\doteq-d(\xi\cdot\mathbb{A})$. Therefore the identity \eqref{eq:smarr:id} becomes a pure surface integral
\be
\label{eq:smarr:A}
0\doteq\int_{\partial\Sigma}\mathbb{Q}[\xi]-\xi\cdot\mathbb{A}
\ee
as we meant to show.
\subsection{The role of scale invariance}
\label{subsec:scale}
In this subsection we show that Eq.\,\eqref{eq:ae:Lshell} is a consequence of a deeper property of \Ae--theory, namely that it is scale invariant. Consider the following constant dilatations of the fields:
\begin{subequations}
\begin{align}
&g_{ab}\to\Omega\,g_{ab},\\
&u^a\to\Omega^{-1/2}\,u^a,\\
&\lambda\to\Omega^{-1}\,\lambda,
\end{align}
\end{subequations}
with $\Omega$ a constant. Since the Christoffel symbol and the Riemann tensor are unaffected by this transformation, and since $\sqrt{-g}\to\Omega^{D/2}\,\sqrt{-g}$, the \Ae\, Lagrangian $D$--form transforms as
\be
\mathbb{L}\to\Omega^{D-2/2}\,\mathbb{L},
\ee
i.e. it experiences a constant rescaling itself: the theory is thus scale invariant (this was already noted in \cite{foster_redefinitions}, where the parameter $A$ there in Eq. (3) corresponds to our $\Omega$). Now consider the corresponding infinitesimal dilatation around the identity, $\Omega\approx1+\omega$,
\be
\label{eq:varl:omega}
\left.
\begin{aligned}
&\delta_\omega g_{ab}=\omega\,g_{ab}\\
&\delta_\omega u^a=-\frac{\omega}{2}u^a\\
&\delta_\omega\lambda=-\omega\lambda
\end{aligned}
\right\}
\implies\delta\mathbb{L}=\omega\left(\frac{D-2}{2}\right)\mathbb{L}.
\ee
From \eqref{eq:varl} and \eqref{eq:ae:theta}
\be
\delta\mathbb{L}\doteq d\Theta(\f,\delta_\omega\f)=\omega\frac{d\left[\left(Y\indices{^m_a}u^a-g^{ab}X\indices{^m_{ab}}\right)\epsilon_m\right]}{2\varkappa},
\ee
and using \eqref{eq:varl:omega} we finally find
\be
\label{eq:onshell:l}
\mathbb{L}\doteq\left(\frac{2}{D-2}\right)\frac{d\left[\left(Y\indices{^m_a}u^a-g^{ab}X\indices{^m_{ab}}\right)\epsilon_m\right]}{2\varkappa},
\ee
in agreement with \eqref{eq:ae:Lshell}.

Accidentally, let us notice that for $D=2$ scale invariance would be an exact symmetry of the Lagrangian, as it is clear form \eqref{eq:varl:omega}, and this would signal the appearance of a conserved charge in the $(1+1)$-dimensional version of the theory.

As a complement to our derivation, let us discuss the inclusion of a cosmological constant. The Lagrangian gets modified by the addition of a term proportional to the volume element
\be
\mathbb{L}=\mathbb{L}_\Ae+\mathbb{L}_\Lambda,\qquad\mathbb{L}_\Lambda=-\frac{\Lambda\epsilon}{\varkappa}.
\ee
This Lagrangian is no more scale invariant; indeed, under the action of the infinitesimal transformation \eqref{eq:varl:omega}, we have
\be
\begin{split}
\delta\mathbb{L}&=\omega\left(\frac{D-2}{2}\right)\mathbb{L}_\Ae+\omega\frac{D}{2}\mathbb{L}_\Lambda\\
&=\omega\left(\frac{D-2}{2}\right)\mathbb{L}+\omega\mathbb{L}_\Lambda.
\end{split}
\ee
Therefore, since the form $\Theta$ is not modified, 
\be
\mathbb{L}\doteq d\mathbb{A}-\left(\frac{2}{D-2}\right)\mathbb{L}_\Lambda
\ee
with the same $\mathbb{A}$ as before, and Eq.\,\eqref{eq:smarr:id} becomes
\be
\label{eq:smarr:lambda}
\begin{split}
0&\doteq\int_{\partial\Sigma}\left[\mathbb{Q}[\xi]-\xi\cdot\mathbb{A}\right]-\left(\frac{2}{D-2}\right)\int_\Sigma\xi\cdot\mathbb{L}_\Lambda\\
&=\int_{\partial\Sigma}\left[\mathbb{Q}[\xi]-\xi\cdot\mathbb{A}\right]+\frac{1}{\varkappa}\left(\frac{2\Lambda}{D-2}\right)\int_\Sigma\xi\cdot\epsilon.
\end{split}
\ee
The last term is only apparently a volume integral. Indeed, as pointed out in \cite{kastor, bazanski}, the Killing equation $\lie_\xi g_{ab}\doteq0$ gives $\nabla_a\xi^a=0$, which in turn implies the existence of an antisymmetric tensor $\xi^{ab}$ such that $\xi^a=\nabla_b\xi^{ab}$. Therefore, using the identity
\be
\label{eq:dnabla}
d\left(W^{a_1\dots a_n}\epsilon_{a_1\dots a_n}\right)=n\left(\nabla_b W^{a_1\dots a_{n-1}b}\right)\epsilon_{a_1\dots a_{n-1}},
\ee
it follows that $\xi\cdot\epsilon\doteq d(\xi^{ab}\epsilon_{ab})/2$, and the volume integral can be turned into a surface one by the Gauss Theorem. So we see that the inclusion of a cosmological constant doesn't invalidate our conclusion that \eqref{eq:smarr:id} is a pure surface integral.

It is clear from our derivation that diffeoinvariant Lagrangians, that are also scale invariant, are total derivatives on shell. Therefore these theories are very good candidates to possess a general Smarr Formula, because the only other requirement that we ask is $\lie_\xi\f\doteq0$, which is generically satisfied along the generators of the horizon \cite{waldentropy1}. 

This should be contrasted with the general case, in which you are not necessarily able to cast \eqref{eq:smarr:id} as a purely surface integral, and therefore to provide a general expression for the Smarr Formula \emph{independently} of the solution. For example in \cite{pacilio_smarr} we derived a Smarr Formula for static Black Holes in Lovelock theory, using Eq. \eqref{eq:smarr:id} \emph{and} a theorem which restricts the general form of the static solutions.
\subsection{Reduction to the static case}
\label{subsec:static}
The Smarr Formula for static, 4--dimensional, spherically symmetric, asymptotically flat Black Holes was studied in \cite{mattingly_mechanics}. The authors impose  staticity from the very beginning, and find that the Smarr Formula eventually follows from a divergence--free antisymmetric two--tensor, $\nabla_b\mathcal{F}^{ab}\doteq 0$. Denoting $s^a$ the spherically symmetric, unit spacelike, vector normal to $u_a$, the tensor $\mathcal{F}^{ab}$ is (see Eq.s (33) and (34) in \cite{mattingly_mechanics})
\be
\label{eq:f}
\mathcal{F}^{ab}=q(u_as_b-s_au_b),
\ee
with
\be
\label{eq:q}
\begin{split}
q=&-\left(1-\frac{c_a}{2}\right)\ut\as+(1-c_\sigma)\st(s^as^b\nabla_au_b)\\ 
&+\frac{c_{123}}{2}\st(\nabla\cdot u),
\end{split}
\ee
and $t^a$ is the Killing field associated to the time--translational symmetries.
Indeed our formalism allows to re-derive, and interpret, this results in a natural way. 

Let us start by noticing that the nullity of the divergence of an antisymmetric two--tensor, $\nabla_b\mathcal{F}^{ab}\doteq 0$, is equivalent to the exactness of a $(D-2)$--form, $d\left(\mathcal{F}^{ab}\epsilon_{ab}\right)\doteq 0$. Now, since Eq. \eqref{eq:smarr:id} holds for any $\Sigma$, we must have $d(\mathbb{Q}[\xi]-\xi\cdot\mathbb{A})\doteq 0$; it is natural to conjecture that $\mathcal{F}^{ab}\epsilon_{ab}$ must be proportional to $\mathbb{Q}[t]-t\cdot\mathbb{A}$, once the latter is restricted to the spherically symmetric case.

First of all, observe that, using Eq.s \eqref{eq:ae:k}, \eqref{eq:ae:yx}, \eqref{eq:ae:q} and \eqref{eq:ae:A},
\begin{multline}
\label{eq:ae:qa:d}
\mathbb{Q}[\xi]-\xi\cdot\mathbb{A}=-\frac{1}{2\varkappa}\left[\nabla^a\xi^b+2c_{13}u^a\xi_c\nabla^{(b}u^{c)}+\right.\\
-\frac{2}{D-2}\left(c_{123}(\nabla\cdot u)u^a\xi^b+c_{14}\xi^a a^b\right)\\
\left.+(\xi\cdot u)(c_1-c_3)\nabla^a u^b-2c_4(\xi\cdot u)u^a a^b\right]\epsilon_{ab},
\end{multline}
or, using the fluid coefficients \eqref{eq:fluid},
\begin{multline}
\label{eq:ae:qa2}
\mathbb{Q}[\xi]-\xi\cdot\mathbb{A}=-\frac{1}{2\varkappa}\left[\nabla^a\xi^b+2c_\sigma u^a\xi_c\nabla^{(b}u^{c)}+\right.\\ 
-\frac{2}{D-2}\left(c_{123}(\nabla\cdot u)u^a\xi^b-c_a\xi^a a^b\right)+c_\sigma(\xi\cdot u)u^a a^b\\
\left.-2c_a(\xi\cdot u)u^a a^b+(\xi\cdot u)c_\omega\,\omega^{ab}\right]\epsilon_{ab},
\end{multline}
where $\omega^{ab}$ is the previously introduced twist of $u^a$.

For a spherically symmetric configuration this expression is simplified. First, the twist $\omega^{ab}$ is null because spherical symmetry implies hypersurface orthogonality. Second, the binormal to the spherically symmetric sections is $\hat{\epsilon}_{ab}=-u_{[a}s_{b]}$. Using $\epsilon_{ab}=\hat{\epsilon}_{ab}\bar{\epsilon}$, with $\bar{\epsilon}$ the volume element of the spherical sections, we find:
\begin{multline}
\label{eq:ae:qa:spherical}
\mathbb{Q}[\xi]-\xi\cdot\mathbb{A}=-\frac{1}{2\varkappa}\left[\nabla^a\xi^b\hat{\epsilon}_{ab}+2c_\sigma\sx(s^as^b\nabla_au_b)\right.\\
\left.-\frac{2c_{123}}{D-2}\sx(\nabla\cdot u)-\frac{2(D-3)c_a}{D-2}\ux\as\right]\bar{\epsilon}.
\end{multline}
Specifying to D=4, we have:
\begin{multline}
\label{eq:ae:qa:spherical2}
\mathbb{Q}[\xi]-\xi\cdot\mathbb{A}=\frac{1}{\varkappa}\left[-\left(1-\frac{c_a}{2}\right)\ux\as\right.\\
\left.+(1-c_\sigma)\sx(s^as^b\nabla_au_b)+\frac{c_{123}}{2}\sx(\nabla\cdot u)\right]\bar{\epsilon}.
\end{multline}
Choosing $\xi^a=t^a$ and comparing \eqref{eq:ae:qa:spherical2} with Eq.s \eqref{eq:f}--\eqref{eq:q}, we see that $\mathcal{F}^{ab}\epsilon_{ab}$ is really proportional to $\mathbb{Q}[t]-t\cdot\mathbb{A}$, as we wanted to show.

Finally, in order to obtain the Smarr Formula, one simply integrates the identity $d(\mathbb{Q}[t]-t\cdot\mathbb{A})\doteq0$ over an hypersurface $\Sigma$ with boundaries $\partial\Sigma=S_\infty\cup S_\text{BH}$.
\section{Extension to IR Ho{\v r}ava gravity}
\label{sec:Ho{\v r}ava}
In \cite{jacobson_Horava} a variant of \Ae--theory was introduced, in which the aether is imposed to be hypersurface othogonal \emph{at the level of the action}: 
\be
\label{eq:ut}
u_a=-N\nabla_aT,\qquad N=(-\nabla_aT\,\nabla_bT\,g^{ab})^{-1/2}.
\ee
The unit--timelike constraint is implicit in the definition \eqref{eq:ut}, and the resulting action is
\be
\label{eq:st}
S_\text{T}=\frac{1}{2\varkappa}\int d^Dx\sqrt{-g}\left[R+L_u\right],
\ee
with $L_u$ the same as before, and where the dynamical variables are now $g^{ab}$ and the ``aether time'' $T$ (from now on in this section we regard $u_a$ as a function of $g_{ab}$ and $T$).

It was shown in \cite{jacobson_Horava} that the action \eqref{eq:st} is equivalent to the IR limit of Ho{\v r}ava gravity, that was originally presented in a canonical form. \Ae--theory \eqref{eq:ae:action2} and IR Ho{\v r}ava gravity \eqref{eq:st} are also related by the fact that an hypersurface orthogonal solution of the former is always a solution of the latter \cite{jacobson_Horava}; moreover they share all the static, spherically symmetric, 4--dimensional BH solutions \cite{uh_maxsym}. Therefore a discussion of the IR Ho{\v r}ava theory case necessarily parallels that of \Ae--theory.

The EOM of \eqref{eq:st} are
\begin{subequations}
\begin{align}
&\nabla_a(N\underleftarrow{\Ae^a})=0 ,\label{eq:st:eom:1}\\
\begin{split}
&G_{ab}=c_1(\nabla_au_m\nabla_bu^m-\nabla_mu_a\nabla^mu_b)+c_4a_aa_b\\
&+\nabla_mX\indices{^m_{ab}}-(u\cdot\Ae) u_au_b-2\Ae_{(a}u_{b)}+\frac{1}{2}L_u\,g_{ab},
\end{split}\label{eq:st:eom:2}
\end{align}
\end{subequations}
where the underleft arrow now denotes projection onto the constant $T$ hypersurfaces. 

The symplectic potential $\Theta$ is
\begin{multline}
\label{eq:st:theta}
\Theta(\f,\delta\f)=\frac{1}{2\varkappa}\left[g_{ab}\nabla^m\delta g^{ab}-\nabla_a\delta g^{ma}+X\indices{^m_{ab}}\delta g^{ab}\right.\\
-2Y\indices{^m_a}u_b\,\delta g^{ab}-Y\indices{^m_c}u^cu_au_b\,\delta g^{ab}\\
\left.+2NY\indices{^m_a}\underleftarrow{\nabla^a\delta T}-2N\underleftarrow{\Ae}^m\delta T\right]\epsilon_m.
\end{multline}
Also the action \eqref{eq:st} is scale invariant, under the scale transformation
\be
\left.
\begin{aligned}
&g_{ab}\to\Omega\,g_{ab}\\
&T\to\Omega^{1/2}\,T
\end{aligned}
\right\}
\implies\mathbb{L}\to\Omega^{D-2/2}\,\mathbb{L}.
\ee
Therefore, by repeating the same argument of Sec.\,\ref{subsec:scale}, we obtain
\be
\mathbb{L}\doteq\left(\frac{2}{D-2}\right)\frac{d\left[\left(Y\indices{^m_a}u^a-g^{ab}X\indices{^m_{ab}}-N\underleftarrow{\Ae}^m\,T\right)\epsilon_m\right]}{2\varkappa}.
\ee
Observe that the last term in the square brackets doesn't contribute on shell, because
\be
\begin{split}
&d[(N\underleftarrow{\Ae}^m\,T)\epsilon_m]
=\nabla_m[N\underleftarrow{\Ae}^m\,T]\epsilon\\
&=[\nabla_m(N\underleftarrow{\Ae}^m)T+N\underleftarrow{\Ae}^m\nabla_mT]\epsilon\\
&\doteq 0,
\end{split}
\ee
and so the on--shell Lagrangian is identical to \eqref{eq:onshell:l}:
\be
\mathbb{L}\doteq\left(\frac{2}{D-2}\right)\frac{d\left[\left(Y\indices{^m_a}u^a-g^{ab}X\indices{^m_{ab}}\right)\epsilon_m\right]}{2\varkappa}.
\ee
However the requirement that $\lie_\xi T\doteq0$ will be not satisfied in many situations of interest. For example, if we consider static, 4--dimensional, spherically symmetric, asymptotically flat BH solutions \cite{barausse_black}, the Smarr Formula is associated with the Killing vector $t^a\equiv(1,0,0,0)$; but since at infinity the aether vector $u^a$ is aligned with $t^a$, it follows that $T$ does depend on $t$ at least in a neighbourhood of spatial infinity; moreover, this in true everywhere in the exact solutions presented in \cite{mattingly_mechanics}. What \emph{is} invariant under the flow of $t^a$ in these solutions is the aether vector $u^a$.

The condition $\lie_\xi u_a\doteq0$ corresponds to the preservation of the foliation. Therefore, in a theory with a preferred foliation, it seems natural to consider this requirement as part of the definition of a Killing field. Let us thus require $\lie_\xi g^{ab}\doteq0$ and $\lie_\xi u_a\doteq0$. Then
\be
\label{eq:st:theta2}
\Theta(\f,\lie_\xi\f)\doteq\frac{\ux\underleftarrow{\Ae}^m\epsilon_m}{\varkappa}.
\ee
\begin{proof}
Since $\lie_\xi g^{ab}\doteq0$, we care only about the last two terms in \eqref{eq:st:theta}. Now $\lie_\xi T=\xi^b\nabla_bT=-\ux/N$. Therefore
\begin{multline}
\varkappa\,\Theta(\f,\lie_\xi\f)=\left[-Y\indices{^m^a}\underleftarrow{\nabla_a\ux}+\ux Y\indices{^m^a}\underleftarrow{\nabla_a\log N}\right.\\
\left.+\underleftarrow{\Ae}^m\ux\right]\epsilon_m=\\
=\left[Y\indices{^m^a}\left(\ux a_a-\underleftarrow{\nabla_a\ux}\right)+\underleftarrow{\Ae}^m\ux\right]\epsilon_m,
\end{multline}
where in the last line we used $a_a=u^b\nabla_b u_a=\underleftarrow{\nabla_a\log N}$. Finally, it can be shown that $\ux a_a-\underleftarrow{\nabla_a\ux}=-\underleftarrow{\lie_\xi u_a}$. Hence, from $\lie_\xi u_a\doteq0$, Eq. \eqref{eq:st:theta2} follows.
\end{proof}
Moreover $\lie_\xi u_a\doteq0$ also implies $\lie_\xi\mathbb{A}\doteq0$. Therefore we see from \eqref{eq:j} that
\be
\label{eq:ae:smarr:correct}
d(\mathbb{Q}[\xi]-\xi\cdot\mathbb{A})\doteq\frac{\ux\underleftarrow{\Ae}^m\epsilon_m}{\varkappa}
\ee
and again the Smarr Formula is obtained by integrating \eqref{eq:ae:smarr:correct} over an appropriate hyeprsurface $\Sigma$. We stress that the Noether form $\mathbb{Q}[\xi]$ for the action \eqref{eq:st} is identical to the one of \Ae--theory \eqref{eq:ae:q}, modulo regarding $u^a$ as a function of $T$.

We conclude this section with some final considerations. If you integrate \eqref{eq:ae:smarr:correct} over a slice of the preferred foliation, the term on the r.h.s. doesn't contribute. The same happens if the solution is also a solution of \Ae--theory, because \eqref{eq:ae:eom:5} holds. Finally the inclusion of a cosmological constant goes exactly as in Sec.\,\ref{subsec:scale}.
\section{Smarr Formula for 3d rotating Black Holes}
\label{sec:rotating}
As we pointed out, the Smarr Formula \eqref{eq:smarr:A} was discussed in connection with the mechanics of Universal Horizons. The previous derivation \cite{mattingly_mechanics, bhatta_thesis} relied upon the assumption of staticity. Our work extends these results, in that expression \eqref{eq:smarr:A} can be applied also to solutions with non--zero angular momentum. 


Slowly rotating solutions have been extensively covered both in the IR Ho{\v r}ava case \cite{wang_slowly, barausse_slowly,sotiriou_uh_review} and the \Ae\, case \cite{barausse_slowly_2}. However, since they are linear in the angular momentum, and since deviations from the static case appear at the quadratic order, we wouldn't gain any clue from them.

On the other hand fully rotating solutions have been found in three spacetime dimensions \cite{btz_uh}, restricted to the parameter space $c_{14}=0$. Although they are not physical, they constitute an arena where to test the effects of rotation. 

In this section, as an application of our work, we derive a Smarr Fomula for these solutions. We remain open minded, and we present the SF both at the Universal and at the Killing Horizon. 
\subsection{The solutions}
In \cite{btz_uh} 3--dimensional exact fully rotating BH solutions of IR Ho{\v r}ava theory were found in the parameter subspace $c_{14}=0$, and assuming a non-null cosmological constant $\Lambda$. They are also solutions of \Ae--theory: to proove this, it is sufficient to verify that the aether EOM\,\eqref{eq:ae:eom:5} is satisfied, because then the metric EOM\,\eqref{eq:ae:eom:6} and \eqref{eq:st:eom:2} are automatically equivalent \cite{jacobson_Horava}. We explicitely checked that $\underleftarrow{\Ae}_a=0$.  In particular this means that the r.h.s. of \eqref{eq:ae:smarr:correct} vanishes.

The solutions are of the form
\begin{subequations}
\label{eq:form:3d}
\begin{align}
&ds^2=-e(r)dt^2+\frac{dr^2}{e(r)}+r^2\left(d\phi^2+\Omega(r)dt^2\right),\\
&u_a=\left(\ut,-\frac{\st}{e(r)},0\right)\label{eq:u:3d},
\end{align}
\end{subequations}
where $t^a=(1,0,0)$ is the time--translational Killing field, and the constraint $u^2=-1$ imposes the consistency relation $\ut^2-\st^2=e(r)$.

We also introduce the rotational Killing field $\phi^a=(0,0,1)$, its normalization $\hat{\phi}^a=(0,0,1/r)$, and the spacelike unit--normal to $u_a$ and $\phi^a$:
\be
\label{eq:3d:s}
s^a=\left(-\frac{\st}{e(r)},-\ut,0\right).
\ee
The Killing vectors $t^a$ and $\phi^a$ are not linearly independent, and they are related by
\be
\label{eq:3d:t}
t^a=-\ut u^a+\st s^a+\Omega(r)\phi^a.
\ee
The set of vectors $(u^a, s^a, \hat{\phi}^a)$ forms an orthonormal triad:
\be
g_{ab}=-u_au_b+s_as_b+\hat{\phi}_a\hat{\phi}_b
\ee
from which one can see that the acceleration vector $a_a=u^b\nabla_bu_a$ is parallel to $s_a$, i.e. $a_a=\as s_a$. This is because $(u\cdot\phi)=0$ globally and $\phi^a$ is a Killing field: therefore $(a\cdot\phi)=\phi^au^b\nabla_bu_a=u^b\nabla_b(u\cdot\phi)-u^au^b\nabla_b\phi_a=0$.

This is the general form of the solutions. The specific expressions for metric functions $e(r)$ and $\Omega(r)$ are
\begin{subequations}
\begin{align}
&e(r)=-\mathcal{M}+\frac{\bar{\J}^2}{4r^2}-\bar{\Lambda}r^2,\\
&\Omega(r)=-\frac{\J}{2r^2},
\end{align}
\end{subequations}
where
\begin{subequations}
\begin{align}
&\bar{\Lambda}=\Lambda-b^2(c_{13}+2c_2),\\
&\bar{\J}^2=(1-c_{13})\J^2-4c_{13}\,a^2,
\end{align}
\end{subequations}
and $\M$, $\J$, $a$ and $b$ are integration constants.

The aether functions $\st$ and $\ut$ are respectively
\begin{subequations}
\begin{align}
&\st=br+\frac{a}{r},\\
&\ut=-\sqrt{e(r)+\left(br+\frac{a}{r}\right)^2}.
\end{align}
\end{subequations}
The parameter $b$ determines the vacuum background of the solutions: the effective cosmological constant $\bar{\Lambda}$ can be positive, negative or null, depending on the value of $b$. Therefore, differently from GR, three--dimensional BH solutions admit also flat and dS backgrounds, in addition to the AdS one.

Observe that $b$ also determines the misalignment at spatial infinity between the aether vector $u^a$ and the vector $t^a$; in particular if $b=0$ then $u^a$ is completely aligned with $t^a$ at spatial infinity.

From now on we restrict, for simplicity, to the case $b=0$.

Before continuing the description of the solutions, we find appropriate to summarize the basic notions of causality behind the definition of the Universal Horizon. As explained in \cite{sotiriou_uh_global}, a curve in the spacetime is said to be causal iff its tangent vector points forward w.r.t. to the aether vector $u^a$. Therefore the causal cones degenerate into local hypersurfaces orthogonal to $u^a$. Now we see that, if $\ut=0$ over an hypersurface, the aether vector points radially and we must distinguish two subcases: (i) when $u^a$ points inward, i.e. in the verse of decreasing $r$, then the configuration represents a Black Hole, because no causal curve can reach spatial infinity; (ii) if $u^a$ points outward in the direction of increasing $r$ then the configuration represents a White Hole, because every causal curve reaches spatial infinity.

Notice that, in the specific case \eqref{eq:form:3d}, it is the sign of $\st_{\text{UH}}$ that determines the nature of the UH: if $\st_{\text{UH}}>0$, the solution describes a Black Hole; viceversa, if $\st_{\text{UH}}<0$, the solution describes a White Hole. When $b=0$, these restrictions on the sign of $\st_\text{UH}$ translate in the sign of the integration constant $a$.

The requirements that a UH exists, and that the foliation is well defined all the way from the center to infinity, imposes restictions on the value of the parameter $a$; in particular, as shown in \cite{btz_uh}, $a$ must satisfy the relation
\be
\label{eq:rel:a}
(1-c_{13})(\J^2+4a^2)=-\frac{\M^2}{\Lambda},
\ee
where we already implemented the restriction $b=0$. We assume both the requirements, and therefore from now on we will view $a$ as expressed in terms of the other parameters by Eq.\,\eqref{eq:rel:a}.  Consequently the location of the Universal Horizon is:
\be
\label{eq:ru}
\ru^2=-\frac{\M}{2\Lambda}.
\ee
Finally the location of the (outer) Killing Horizon, defined by $e(r)\overset{\text{KH}}{=}0$, is
\be
\label{eq:rk}
\rk^2=-\frac{\M}{2\Lambda}\left[1+\sqrt{1+\frac{\Lambda\bar{\J}^2}{\M^2}}\right],
\ee
from which we immediately see that $\rk>\ru$.
\subsection{Mass and angular momentum}
\label{sec:mass}
The aim of this section is to compute the mass and the angular momentum of the 3--dimensional solution.
 
Since the Smarr Formula for a $D$--dimensional BH has the form
\be
\label{eq:smarr:d}
\left(D-3\right)M=\left(D-2\right)TS+\left(D-2\right)\Omega J+\dots
\ee
we don't need to know the exact expression of the mass $M$ when $D=3$. However, we will compute it anyway for pedagogical reasons, because it represents the first computation of the mass in Ho{\v r}ava gravity and \Ae--theory for an AdS asymptotic infinity, and to illustrate the covariant Hamiltonian subtraction scheme.

The Hamiltonian analysis of Ho{\v r}ava theory was carried out in \cite{Horava_hamiltonian}, where an expression for the energy of asymptotically flat (AF) configurations was also derived. It agrees with the energy for AF spacetimes in \Ae--theory, as derived both in \cite{foster_noether} using the covariant Hamiltonian formalism a la Wald, and in \cite{eling_energy} using pseudotensor methods. Moreover in \cite{jacobson_positive} a positivity theorem for such AF energy was proven. Therefore the notion of energy in the AF case is well established. 

However we want to deal with the more general asymptotics of the previous solutions. We follow a different strategy than those followed in the AF case. It consists in a background subtraction procedure \cite{regge_teitelboim,hawking_horowitz, brown_york}, using the covariant Hamiltonian formalism \cite{waldentropy1, waldentropy2}, in the form that we already used in \cite{pacilio_smarr} to compute the mass of static Lovelock Black Holes. This procedure can be seen as a covariantization of the Regge--Teitelboim one.

Our starting point is the observation that, in general, the background carries a non null total energy: therefore the total mass of the BH is the excess of total energy w.r.t. to the background. At the same time, at spatial infinity, a BH solution can be viewed as a linear perturbation around the asymptotic background, and \emph{there is} an expression for the linear variation of the Hamiltonian w.r.t. a given background solution: as shown in  \cite{waldentropy1, waldentropy2}, the Hamilton equations relative to the flow of a generic vector field $\xi$ over an initial value surface $\Sigma$ are given by
\be
\label{eq:deltah}
\delta H[\xi]=\int_{\partial\Sigma}\left[\delta\mathbb{Q}[\xi]-\xi\cdot\Theta(\f_0,\delta\f)\right]
\ee
where $\f_0$ indicates the background fields and $\delta\f$ their variations.

It is then natural to identify the variations of the energy as \cite{waldentropy1, waldentropy2}
\be
\label{eq:dele}
\delta\mathcal{E}=\int_{S_\infty}\left[\delta\mathbb{Q}[t]-t\cdot\Theta(\f_0,\delta\f)\right],
\ee
where $S_\infty$ is the outer boundary of $\partial\Sigma$, and $t^a$ is the time--translational Killing field at infinity. 

We adopt Eq.\,\eqref{eq:dele} as our definition of BH mass. For simplicity we restrict to the case $b=0$. 

From \eqref{eq:u:3d} the aether time $T$ is
\be
\label{eq:T}
T=t-\int^rdr'\frac{\st}{e(r)\ut},
\ee
which in the case $b=0$ reduces to
\be
\label{eq:T2}
T=t-\frac{1}{3}\frac{a}{\left(\Lambda\right)^{3/2}r^3}+O\left(\frac{1}{r^5}\right).
\ee
Therefore the partition between background fields and perturbations at asymptotic infinity is:
\be
\label{eq:phi0}
\f_0=
\left\{
\begin{aligned}
&e_0(r)=-\Lambda r^2\\
&\Omega_0(r)=0\\
&T_0=t
\end{aligned}
\right.,
\ee
\be
\label{eq:delphi}
\delta\f=
\left\{
\begin{aligned}
&\delta e(r)=-\M+\frac{\bar{\J}^2}{4r^2}\\
&\delta\Omega(r)=\Omega(r)=-\frac{\J}{2r^2}\\
&\delta T=-\frac{1}{3}\frac{a}{\left(\Lambda\right)^{3/2}r^3}+O\left(\frac{1}{r^5}\right)
\end{aligned}
\right..
\ee
From \eqref{eq:phi0} we see that $\f_0$ describes an AdS solution, which is obtained by taking the limit $\M,\J\to0$ of the full solution, as it is clear from \eqref{eq:delphi} (recall that we regard $a$ as a function of $\M$ and $\J$).

We are now ready to apply \eqref{eq:dele} to compute the total mass. We computed the relevant integrals with Mathematica, and here we just spell out the main steps. First of all
\be
\int_{S_\infty}\mathbb{Q}[t]=-\lim_{r\to\infty}\frac{\Lambda r^2}{4G}+O\left(\frac{1}{r^2}\right),
\ee
from which we see that
\be
\label{eq:step1}
\int_{S_\infty}\delta\mathbb{Q}[t]=O\left(\frac{1}{r^2}\right).
\ee
Second, the last term in \eqref{eq:st:theta} doesn't contribute, because the solution satisfies also the \Ae\, EOM $\underleftarrow{\Ae}_a=0$. Moreover, since the binormal to the circular sections is $\hat{\epsilon}_{ab}=-2u_{[a}s_{b]}=-2\delta^t_{[a}\delta^r_{b]}$, then
\be
\int_{S_\infty}-t\cdot\Theta=\int_{S_\infty}\Theta^r\bar{\epsilon}
\ee
where we have used the notation $\Theta=\Theta^m\epsilon_m$. The result is
\be
\label{eq:step2}
\int_{S_\infty}-t\cdot\Theta=\frac{\M}{8G}.
\ee
Putting together Eq.s \eqref{eq:step1} and \eqref{eq:step2} the total mass is
\be
\label{eq:M}
M=\frac{\M}{8G}.
\ee
The computation of the angular momentum $J$ in much easier: as shown in \cite{waldentropy1, waldentropy2} the total angular momentum is just
\be
J=-\int_{S_\infty}\mathbb{Q}[\phi],
\ee
where $\phi^a$ is the rotational Killing field. In the present case the result is
\be
\label{eq:J}
J=\left({1-c_{13}}\right)\frac{\J}{8G}.
\ee
%
\subsection{The Smarr Formula}
\label{sec:3d:smarr}
\subsubsection{Smarr Formula at the Killing Horizon}
The Smarr Formula at the Killing Horizon is given by Eq.\,\eqref{eq:smarr:lambda} for $D=3$:
\be
\label{eq:3d:smarr1}
0=\int_{\partial\Sigma}\left[\mathbb{Q}[\xi]-\xi\cdot\mathbb{A}\right]+\frac{2\Lambda}{\varkappa}\int_\Sigma\xi\cdot\epsilon,
\ee
where $\Sigma$ is an hypersurface with boundaries $C_\infty$ and $C_{\text{KH}}$, respectively the circular sections at asymptotic infinity and at the Killing Horizon.

The vector field $\xi^a$ must be chosen as the Killing field generating the KH:
\be
\label{eq:x:kil}
\xi^a=t^a+\Omega_\text{KH}\phi^a,\quad\Omega_\text{KH}=-\Omega(\rk),
\ee
where $\Omega_{\text{KH}}$ is the frame--dragging angular velocity at the KH. 

Having identified $\xi^a$, we proceed to evaluate all the terms in \eqref{eq:3d:smarr1}. The second term is simply
\be
\label{eq:smarr:step1}
\frac{2\Lambda}{\varkappa}\int_\Sigma\xi\cdot\epsilon=\lim_{r\to\infty}\frac{\Lambda r^2}{4G}-\frac{\Lambda\,\rk^2}{4G}.
\ee
The divergent part is compensated by the first part in
\be
\label{eq:smarr:step2}
\int_{C_\infty}\left[\mathbb{Q}[\xi]-\xi\cdot\mathbb{A}\right]=-\lim_{r\to\infty}\frac{\Lambda r^2}{4G}-\Omega_\text{KH}\J.
\ee
Finally, as shown in Appendix \ref{sec:app1}, 
\be
\label{eq:smarr:step3}
\int_{C_\text{KH}}\left[\mathbb{Q}[\xi]-\xi\cdot\mathbb{A}\right]=\left(\frac{\kk}{2\pi}\right)\frac{P_\text{KH}}{4G}-\frac{c_{13}}{4G}\frac{a^2}{\rk^2},
\ee
where $\kk$ and $P_\text{KH}$ are, respectively, the surface gravity and the perimeter of the KH.

Therefore, putting togheter Eq.s\,\eqref{eq:smarr:step1}--\eqref{eq:smarr:step3}, the Smarr Formula at the KH becomes
\be
\label{eq:3d:smarr3}
0\cdot M=\left(\frac{\kk}{2\pi}\right)\frac{P_\text{KH}}{4G}+\Omega_\text{KH}\J+\frac{\Lambda\,\rk^2}{4G}-\frac{c_{13}}{4G}\frac{a^2}{\rk^2}.
\ee
\subsubsection{Smarr Formula at the Universal Horizon}
The Smarr Formula at the Universal Horizon is again given by Eq.\,\eqref{eq:3d:smarr1}, where now $\partial\Sigma=C_\infty\cup C_{\text{UH}}$.

There is an ambiguity in the choice of the vector field $\xi$. Indeed the UH is defined by the equation $\ut_{\text{UH}}=0$. However, since $(u\cdot\phi)=0$ globally, any vector field of the form $\xi^a=t^a+\alpha\,\phi^a$, with $\alpha$ generic, satisfies the defining equation $\ux_{\text{UH}}=0$. We proceed in analogy with the KH, and we choose 
\be
\xi^a=t^a+\Omega_{\text{UH}}\phi^a,\quad\Omega_{\text{UH}}=-\Omega(\ru),
\ee 
where again $\Omega_{\text{UH}}$ is the frame--dragging angular velocity at the UH. From \eqref{eq:3d:t}
\be
\xi^a\overset{\text{UH}}{=}-\st s^a,
\ee
and therefore $\xi^a$ is also orthogonal to the circular cross sections of the UH, analogously to what happens at the KH.

Repeating the same steps as before, we have
\be
\label{eq:smarr:step4}
\frac{2\Lambda}{\varkappa}\int_\Sigma\xi\cdot\epsilon=\lim_{r\to\infty}\frac{\Lambda r^2}{4G}-\frac{\Lambda\,\ru^2}{4G}
\ee
and also
\be
\label{eq:smarr:step5}
\int_{C_\infty}\left[\mathbb{Q}[\xi]-\xi\cdot\mathbb{A}\right]=-\lim_{r\to\infty}\frac{\Lambda r^2}{4G}-\Omega_{\text{UH}}\J,
\ee
from which we see the compensation of the divergent parts. Finally, as shown in Appendix \ref{sec:app2},
\be
\label{eq:smarr:step6}
\int_{C_\text{UH}}\left[\mathbb{Q}[\xi]-\xi\cdot\mathbb{A}\right]=\frac{\left(1-c_{13}\right)}{4G}\frac{a^2}{\ru^2}.
\ee
Putting together Eq.s\,\eqref{eq:smarr:step4}--\eqref{eq:smarr:step6}, the Smarr Formula at the UH becomes\footnote{In the limit $\J\to0$ Eq. (4.24) of \cite{btz_charged} is recovered.}
\be
\label{eq:3d:smarr4}
0\cdot M=\frac{\left(1-c_{13}\right)}{4G}\frac{a^2}{\ru^2}+\Omega_{\text{UH}}\J+\frac{\Lambda\,\ru^2}{4G}.
\ee
%
\subsubsection{Thermodynamical interpretation}

In this section we want to comment about the possible thermodynamical interpretation of the two Smarr Formulas \eqref{eq:3d:smarr3} and \eqref{eq:3d:smarr4}. 
We start from the easier case, the one at the Killing Horizon. The first three terms on the r.h.s. of \eqref{eq:3d:smarr3} can be immediately interpreted in the following usual manner:
\begin{itemize}
\item $\kk/2\pi$ is the Hawking temperature of the KH, with associated entropy $S=P_\text{KH}/4G$;
\item $\Omega_\text{KH}\J$ is the work term associated to the presence of a non-zero angular momentum;
\item $\Lambda\rk^2/4G$ is in the form $-2\,pV$, where $p=-\Lambda/8\pi G$ is the pressure of the cosmological fluid, while $V=\pi\rk^2$ is the euclidean volume of the circle bounded by the Killing Horizon. The presence of such a term in the Smarr Formula is typical of Black Holes with a cosmological constant \cite{kastor_ads,lambda_review}.
\end{itemize}
The last term, $-\frac{c_{13}}{4G}\frac{a^2}{\rk^2}$, can be viewed as an additional work term originating from aether degrees of freedom. 
While it  would be beyond the scope of the present paper an attempt to identify these possible degrees of freedom we do think that this could be a relevant stream of investigation worth pursuing in the future.

The Universal Horizon case is a bit more subtle. First of all, guided by the analogy with the KH, we split the first term of \eqref{eq:3d:smarr4} as follows:
\be
\label{eq:3d:smarr5}
0\cdot M=\frac{a^2}{4G\,\ru^2}+\Omega_{\text{UH}}\J+\frac{\Lambda\,\ru^2}{4G}-\frac{c_{13}}{4G}\frac{a^2}{\ru^2},
\ee
in such a way that the last three terms in \eqref{eq:3d:smarr5} have the same interpretation as their Killing counterparts: rotational work term, pressure--volume term of the cosmological fluid, and aether DOF work term. 

What misses is to cast the first term in a $TS$ form. This in turn requires a notion of temperature of the Universal Horizon, a subject that is currently under investigation: let us therefore briefly sum up the status of the work so far.

A candidate expression for the temperature of the UH was first estimated in \cite{mattingly_tunneling} via a tunnelling approach, by computing the tunnelling probability across the UH of a scalar field with a quadratic ultraviolet dispersion relation in the aether preferred frame
\be
\omega\sim p^2,
\ee
such that the modes have the UH as their causal horizon. The authors considered the two specific static BH exact solutions of \cite{mattingly_mechanics}, finding the tunnelling temperature
\be
\label{eq:tuh1}
T_\text{UH}=\frac{\ku}{2\pi},
\ee
where the quantity $\ku$ is defined as
\be
\ku=\left.\frac{1}{2}u^a\nabla_a\ut\right|_\text{UH}.
\ee
The computation was then generalized in \cite{ding_radiation} to more general scaling dispersion relations\footnote{$N=1$ is excluded because it corresponds to relativistic modes.}
\be
\label{eq:omegan}
\omega\sim p^N,\quad N>1,
\ee
finding the tunneling temperature
\be
\label{eq:tuh2}
T_\text{UH}^{(N)}=\left(\frac{N-1}{N}\right)\frac{\ku}{\pi},
\ee
which is consistent with \eqref{eq:tuh1} for $N=2$.

Moreover \cite{cropp_ray, cropp_thesis} studied the peeling behaviour of the modes \eqref{eq:omegan} at the UH, finding that the peeling surface gravity of the UH is
\be
\label{eq:kpeel}
\kappa^{(N)}_\text{peel}=\left(\frac{N-1}{N}\right)\ku.
\ee
Therefore $T_\text{UH}^{(N)}$ is related to the surface gravity of the UH by
\be
T_\text{UH}^{(N)}=\frac{\kappa_\text{peel}^{(N)}}{\pi}.
\ee
(Notice a difference by a factor of 2 w.r.t. the Killing horizon.)

From Eq.\,\eqref{eq:tuh2} it is apparent that $T^{(N)}_\text{UH}$ depends on $N$, i.e. it depends on the specific dispersion relation of the test field. This is disturbing, especially in consideration of the fact that it seems to imply that the (would be) entropy of the UH is species dependent, rather than being universal.

However, this implication implicitly assumes knowledge of UV completion of both gravity and matter sectors as this will be always relevant at the UH. Since \Ae--theory is not UV complete, one cannot be sure that the N dependence of the temperature is necessarily problematic, as indeed the low energy component of the Hawking spectrum at infinity will be anyway heavily influenced by the KH, while the UV part would be undetermined without a UV completion of the gravity and matter sectors. 

Ho{\v r}ava gravity, instead, aspires to be a UV complete gravitational theory, to be eventually supported by a similar completion in the matter sector, therefore we cannot ignore the problem. A possible solution could be the requirement of an extra symmetry, such as invariance of the action under Lifshitz anisotropic scaling, that would suffice to enforce a universal value of $N$.\footnote{We are grateful to D. Mattingly for having pointed this consideration to us.} 

By the way, notice that in Ref. \cite{btz_uh_uv} static and rotating universal horizons in (2+1) Ho{\v r}ava gravity where found, in the version of the theory supplied with an extra $U(1)$ symmetry (see \cite{wang_review} for a review of the various versions of the theory). The solutions are exact up to the ultraviolet regime, therefore opening the possibility to study the dependence of the scaling $N$ in a UV universal horizon background.

Alternatively, given a theory describing interacting fields with modified dispersion relations at the tree level, the dynamics could be such that radiative corrections will equalise the UV behaviour of the dispersion relations. 

Finally, if one embraces an EFT point of view, one can arbitrarily expand the action in inverse powers of a Lorentz breaking scale (assumed to be smaller of the UV cutoff of the EFT), therefore effectively inducing $N\to\infty$ for all the species. 

In conclusion, possible mechanisms that save the universality of (either a finite or infinite) $N$ can be envisaged. In what follows we shall assume that one of these is indeed realized, and hence adopt Eq.\,\eqref{eq:tuh2} as our definition of UH temperature.

For the 3--dimensional solution considered in this section, in the subcase $b=0$, we have
\be
\label{eq:kuh2}
\ku=\frac{a\sqrt{-\Lambda}}{\ru}.
\ee
So we can rewrite the Smarr Formula \eqref{eq:3d:smarr5} as
\be
\label{eq:3d:smarr6}
0\cdot M=T_\text{UH}^{(N)}\left(\frac{\pi\,c_N}{4G}\frac{a}{\sqrt{-\Lambda}\,\ru}\right)+\Omega_{\text{UH}}\J+\frac{\Lambda\,\ru^2}{4G}-\frac{c_{13}}{4G}\frac{a^2}{\ru^2},
\ee
where we defined $c_N=N/(N-1)$, from which we tentatively identify 
\be
\label{eq:3d:entropy}
S=\frac{\pi\,c_N}{4G}\frac{a}{\sqrt{-\Lambda}\,\ru}.
\ee
Notice that $S$ is not proportional to the perimeter $P_\text{UH}$ of the UH in the fully rotating case, while it becomes such in the static case $J=0$:
\be
\lim_{J\to0}S=\frac{c_N\,P_{\text{UH}}}{8G\sqrt{1-c_{13}}}.
\ee
This concludes our thermodynamical interpretation of the UH Smarr Formula \eqref{eq:3d:smarr5}.

It must be noted that the Smarr Formula is just a consistency relation between the mass and the other parameters of the solution. It doesn't say anything about physical processes. In this respect a study of the First Law of mechanics would be enlightening of the nature of the various terms that we have tentatively identified in this section. We reserve this investigation for a future work \cite{preparation}.
\section{Conclusions}
\label{sec:conclusion}
In this paper we presented a general method for deriving the Smarr Formula of \Ae--theory and IR Ho{\v r}ava gravity, which makes crucial use of the scale invariance property of these theories.

Our derivation extends previous results \cite{mattingly_mechanics, bhatta_thesis} that were applicable only to static Black Hole configurations. 
This opens the possibility to study the SF of rotating BHs, in order to gain some insight about their possible thermodynamic behaviour.

As an application we analysed the Smarr Formula for the 3--dimensional rotating BH solutions found in \cite{btz_uh}, in the two cases in which the internal boundary is chosen at the Universal Horizon or at the Killing Horizon. 
We found that both the cases admit a possible thermodynamical interpretation of the terms appearing in the SF, which implies the appearance of extra work terms originating from aether degrees of freedom. The nature of such terms surely deserves further investigation.

The above results are also interesting in view of the still not solved controversy about which of the two horizons is responsible for the thermodynamics \cite{parentani}. More insights in this subject can be obtained through the analysis of First Laws corresponding to the above Smarr Formulas, and we reserve this study to a subsequent paper.

Finally, although 4--dimensional fully rotating BH solutions still don't exist, we believe that, in future perspective, they represent the more promising research line where the techniques presented in this paper can be eventually applied, to have a better understanding of the thermodynamics of Lorentz violating gravitational theories.
%
\begin{acknowledgements}
The authors are grateful to Jishnu Bhattacharyya, D. Mattingly, R. Parentani, T. Sotiriou and  M. Visser for helpful and constructive discussions.
S.L. acknowledges financial support from the John Templeton Foundation (JTF) grant \#51876.
\end{acknowledgements}
\begin{appendices}
\section{Computation of the Smarr Formula at the KH}
\label{sec:app1}
In this Appendix we prove Eq.\,\eqref{eq:smarr:step3}. Recall that we are working in the limit $b=0$.

From Eq.\,\eqref{eq:ae:qa2}, specifying to $D=3$ and to $c_{14}=0$, we have
\begin{multline}
\label{eq:app1:qa1}
\mathbb{Q}[\xi]-\xi\cdot\mathbb{A}=-\frac{1}{2\varkappa}\left[\nabla^a\xi^b+2c_{13} u^a\xi_c\nabla^{(b}u^{c)}\right.\\ 
\left.-2c_{123}(\nabla\cdot u)u^a\xi^b+c_{13}\ux u^aa^b\right]\epsilon_{ab}.
\end{multline}
To manipulate the first term, observe that $\epsilon_{ab}=\hat{\epsilon}_{ab}\bar{\epsilon}$ and, by definition of surface gravity, $\nabla^a\xi^b\hat{\epsilon}_{ab}\overset{\text{KH}}{=}-2\kk$; therefore
\begin{multline}
\label{eq:app1:qa2}
\mathbb{Q}[\xi]-\xi\cdot\mathbb{A}\overset{\text{KH}}{=}\frac{\kk\bar{\epsilon}}{\varkappa}-\frac{1}{2\varkappa}\left[2c_{13} u^a\xi_c\nabla^{(b}u^{c)}\right.\\ 
\left.-2c_{123}(\nabla\cdot u)u^a\xi^b+c_{13}\ux u^aa^b\right]\epsilon_{ab}.
\end{multline}
Then, using $\hat{\epsilon}_{ab}=-2u_{[a}s_{b]}$, we can reduce also the remaining terms:
\begin{multline}
\label{eq:app1:qa3}
\mathbb{Q}[\xi]-\xi\cdot\mathbb{A}\overset{\text{KH}}{=}\frac{\kk\bar{\epsilon}}{\varkappa}-\frac{1}{2\varkappa}\left[2c_{13} s^a\xi^b\nabla_{(a}u_{b)}\right.\\ 
\left.-2c_{123}\sx(\nabla\cdot u)+c_{13}\ux\as\right]\bar{\epsilon}.
\end{multline}
Now, from \eqref{eq:3d:t} and \eqref{eq:x:kil}, $\xi^a=-\ut u^a+\st s^a$, from which it follows
\begin{multline}
\label{eq:app1:qa4}
\mathbb{Q}[\xi]-\xi\cdot\mathbb{A}\overset{\text{KH}}{=}\frac{\kk\bar{\epsilon}}{\varkappa}-\frac{1}{\varkappa}\left[c_{13}\st\left(s^as^b\nabla_a u_b\right)\right.\\ 
\left.-c_{123}\st(\nabla\cdot u)\right]\bar{\epsilon}.
\end{multline}
A direct computations shows that, for $b=0$, 
\be
\label{eq:app1:auxiliary}
(s^a s^b\nabla_a u_b)=\frac{a}{r^2}\quad\text{and}\quad (\nabla\cdot u)=0,
\ee
and, since in the same limit $\st=a/r$,
\be
\label{eq:app1:qa5}
\mathbb{Q}[\xi]-\xi\cdot\mathbb{A}\overset{\text{KH}}{=}\frac{\kk\bar{\epsilon}}{\varkappa}-\frac{c_{13}\,\bar{\epsilon}}{\varkappa}\frac{a^2}{ \rk^3}.
\ee
The only part depending on the coordinate $\phi$ is the circular line element $\bar{\epsilon}=\ru d\phi$. We can finally integrate over $C_{\text{KH}}$, thus obtaining
\be
\label{eq:app2:qa6}
\int_{C_\text{KH}}\left[\mathbb{Q}[\xi]-\xi\cdot\mathbb{A}\right]=\frac{\kk\,P_\text{KH}}{8\pi G}-\frac{c_{13}}{4G}\frac{a^2}{ \rk^2}.\quad\blacksquare
\ee
\section{Computation of the Smarr Formula at the UH}
\label{sec:app2}
In this Appendix we prove Eq.\,\eqref{eq:smarr:step6}.

From Eq.\,\eqref{eq:ae:qa2}, specifying to $D=3$ and to $c_{14}=0$, and using the fact that $\ux_{\text{UH}}=0$, we have
\begin{multline}
\label{eq:app2:qa1}
\mathbb{Q}[\xi]-\xi\cdot\mathbb{A}\overset{\text{UH}}{=}-\frac{1}{2\varkappa}\left[\nabla^a\xi^b+2c_{13} u^a\xi_c\nabla^{(b}u^{c)}\right.\\ 
\left.-2c_{123}(\nabla\cdot u)u^a\xi^b\right]\epsilon_{ab}.
\end{multline}
Using $\epsilon_{ab}=-2u_{[a}s_{b]}\bar{\epsilon}$, it becomes
\begin{multline}
\label{eq:app2:qa2}
\mathbb{Q}[\xi]-\xi\cdot\mathbb{A}\overset{\text{UH}}{=}-\frac{1}{2\varkappa}\left[2s^au^b\nabla_a\xi_b+2c_{13} s^a\xi^b\nabla_{(a}u_{b)}\right.\\ 
\left.-2c_{123}\sx(\nabla\cdot u)\right]\bar{\epsilon},
\end{multline}
where in the first term inside the squared brackets we used the Killing equation $\nabla_a\xi_b=\nabla_{[a}\xi_{b]}$. This term can be further manipulated as follows:
\be
\label{eq:app2:1}
s^au^b\nabla_a\xi_b=s^a\nabla_a\ux-s^a\xi^b\nabla_au_b;
\ee
but $\ux$ depends only on $r$, therefore the first term on the r.h.s of \eqref{eq:app2:1} selects the radial part of $s^a$, which from \eqref{eq:3d:s} is zero at the UH. Therefore $s^au^b\nabla_a\xi_b\overset{\text{UH}}{=}-s^a\xi^b\nabla_au_b$, and \eqref{eq:app2:qa2} becomes
\begin{multline}
\label{eq:app2:qa3}
\mathbb{Q}[\xi]-\xi\cdot\mathbb{A}\overset{\text{UH}}{=}\frac{1}{\varkappa}\left[s^a\xi^b\nabla_a u_b-c_{13} s^a\xi^b\nabla_{(a}u_{b)}+\right.\\ 
\left.+c_{123}\sx(\nabla\cdot u)\right]\bar{\epsilon}.
\end{multline}
Now, using $\xi^a\overset{\text{UH}}{=}\st s^a$, we have
\begin{multline}
\label{eq:app2:qa4}
\mathbb{Q}[\xi]-\xi\cdot\mathbb{A}\overset{\text{UH}}{=}\frac{1}{\varkappa}\left[\left(1-c_{13}\right)\st\left(s^a s^b\nabla_a u_b\right)\right.\\ 
\left.+c_{123}\st(\nabla\cdot u)\right]\bar{\epsilon}.
\end{multline}
Using \eqref{eq:app1:auxiliary} as before we obtain
\be
\label{eq:app1:qa5}
\mathbb{Q}[\xi]-\xi\cdot\mathbb{A}\overset{\text{UH}}{=}\frac{\bar{\epsilon}}{\varkappa}\left(1-c_{13}\right)\frac{a^2}{\ru^3}.
\ee
Finally, integrating over $C_\text{UH}$, we obtain
\be
\label{eq:app1:qa6}
\int_{C_{\text{UH}}}\left[\mathbb{Q}[\xi]-\xi\cdot\mathbb{A}\right]=\frac{\left(1-c_{13}\right)}{4G}\frac{a^2}{\ru^2}.
\ee
thus proving Eq.\,\eqref{eq:smarr:step3}. $\blacksquare$
\end{appendices}
\end{document}